# The Interaction Gap: A Step Toward Understanding Trust in Autonomous Vehicles Between Encounters

Jacob G. Hunter[1], Matthew Konishi[1], Neera Jain[1], Kumar Akash[2],
Xingwei Wu[2], Teruhisa Misu[2], Tahira Reid[1]
[1]Dept. of Mechanical Engineering, Purdue University, Indiana
[2]Honda Research Institute, San Jose, California

Shared autonomous vehicles (SAVs) will be introduced in greater numbers over the coming decade. Due to rapid advances in shared mobility and the slower development of fully autonomous vehicles (AVs), SAVs will likely be deployed before privately-owned AVs. Moreover, existing shared mobility services are transitioning their vehicle fleets toward those with increasingly higher levels of driving automation. Consequently, people who use shared vehicles on an "as needed" basis will have infrequent interactions with automated driving, thereby experiencing *interaction gaps*. Using human trust data of 25 participants, we show that interaction gaps can affect human trust in automated driving. Participants engaged in a simulator study consisting of two interactions separated by a one-week interaction gap. A moderate, inverse correlation was found between the change in trust during the initial interaction and the interaction gap, suggesting people "forget" some of their gained trust or distrust in automation during an interaction gap.

## INTRODUCTION

Autonomous vehicles (AVs) are appealing due to their potential benefits in safety, mobility for the disabled, and on-demand ride services (Fagnant and Kockelman, 2015). Several automakers are currently focusing on the development of shared autonomous vehicles (SAVs) (Narayanan et al., 2020). The expected release of SAVs before privately-owned AVs is a consequence of the high development and production costs of these vehicles, as well as the recent interest and innovations in shared mobility. Although SAVs "may be widely available by the 2030s" (Litman, 2017), due to technical and legislative constraints, not all shared mobility services will immediately provide high or full autonomy, but several services may remain partially automated. Regardless of the level of automation, one of the most important aspects of near- and long-term SAV adoption is calibrating consumer trust in a mode of transport that is used on an "as-needed" basis.

Trust in automation or autonomous vehicles is not a new topic; however, despite three decades of research on this topic (Muir, 1994; Lee and See, 2004; Hoff and Bashir, 2015), few studies have investigated changes in human trust due to multiple distinct interactions with an automated system. In 2013, Beggiato and Krems oversaw a simulator study in which they measured participants' trust in a vehicle's adaptive cruise control (ACC) over three identical trials. Although participants had multiple interactions with the vehicle, the trials were variably spaced (7-24 days) depending upon participants' schedules. Additionally, participants' trust was not measured at the beginning of subsequent trials. The trust measurements, collected before and after trial 1, after trial 2, and after trial 3, allowed the researchers to observe changes in trust over time; however, they were unable to evaluate how trust changed solely during the gaps between trials, thereby losing some of the participants' trust dynamics. In 2015, Beggiato et al. conducted a human subjects study with an on-road vehicle with ACC capabilities. They found that across 10 interactions, trust increased according to a power function, "levelling off after approximately the fifth session." The 10 interactions were spread over two months, but again, trust was only measured after an interaction and not a gap. However, an eight-week interaction gap was studied by Hartwich et al. (2019) in which they measured trust after each participant's final AV simulator drive during their first visit and measured trust again eight weeks later before each participant's drive in a real AV. A repeated measures ANOVA showed no significant difference between trust after the driving simulator trial and trust eight weeks later before the real AV trial. The reliability of the simulated AV and the actual AV was 100%. Currently though, AVs are not 100% reliable. As such, it is critical to understand how trust changes during an interaction gap with an automated system that can perform unreliably.

We designed and conducted a human subjects study to answer these research questions: (i) What happens to trust in a partial, SAE Level 2 AV during an interaction gap of one week? (ii) Which aspects of trust in automation change, and how do they change, during an interaction gap?

The human subjects study was conducted on a medium fidelity driving simulator at Purdue University. Current regulations and technology allow for SAE Level 2 vehicles (partial driving automation) to be on the road. Thus, our use of an SAE Level 2 AV (SAE, 2021) in the driving simulation mirrors current standards while providing an understanding of an AV interaction gap in preparation for SAVs.

## METHODOLOGY

We define an interaction gap as a significant period of time between subsequent uses of an automated system. Interaction gaps vary in time depending upon an operator's needs, preferences, or the availability of the automated system. Interaction gaps should not be confused with "transition gaps." A transition gap is a *brief* period of time between subsequent uses of an automated system. Therefore, one interaction may be comprised of two or more "micro-interactions" joined together by one or more transition gaps.

To put these definitions in perspective with the literature, Hoff and Bashir (2015) present a three-layer trust model where trust in automation is captured by dispositional, situational, and learned trust. Dispositional trust depends upon an individual's demographic factors and personality traits while situational trust is influenced by the specific context of an interaction. Learned trust, however, constitutes "an operator's evaluations of a system drawn from past experience or the current interaction;" thus, learned trust can be classified as a combination of "initial learned trust" and "dynamic learned trust," respectively (Hoff and Bashir, 2015). Thus, initial learned trust is measured after an interaction gap whereas dynamic learned trust is measured over the course of an interaction.

For the purposes of this work, we measure trust at the beginning and end of an interaction gap. We also measure trust once during each transition gap to capture dynamic learned trust. Thus, the interaction and transition gap(s) allow us to measure user trust and alter experimental conditions of the automation for the subsequent (micro-)interaction (see Figure 1).

**Measuring Trust**

Previous research examined drivers' subjective trust assessments of a simulated SAE Level 2 AV that uses augmented reality (AR) graphic cues to communicate with the driver (Wu et al., 2020). However, only one survey question on confidence is used to evaluate participants' trust. In this work, a holistic tool is used to capture participants' cumulative self-reported trust of a simulated SAE L2 AV over varying drive conditions. This tool, a 12-item questionnaire derived from word clusters is specifically designed to evaluate trust between humans and automation (Jian et al., 2000). The advantage of this questionnaire is that it utilizes both positive and negative phrases to capture multiple dimensions of trust. Many researchers average several or all the trust values from this questionnaire to produce a single trust score (Beggiato and Krems, 2013; Gold et al., 2015; Helldin et al., 2013; Verberne et al., 2012); however, one of the original authors of the scale demonstrates the value of observing each question's score individually to understand the different aspects of trust (Bisantz and Seong, 2001). The average and individual trust scores(s) are used in the data analysis in the Results Section.

**Participants**

Thirty-eight participants from the West Lafayette area completed at least a portion of the study. Two participants did not return for the second interaction, one was dismissed for not following instructions, two were excluded for not receiving adequate instruction, and an additional eight were excluded for experiencing simulator glitches unknowingly, which may have affected their trust response. Therefore, data from 25 participants (12 male; 13 female) is used for analysis. The age ranged from 18-72 years ($\bar{x}$ = 26). Participants were paid $15 per hour. Screening criteria were: (i) 18+ years (ii) legally allowed to drive in the U.S. (iii) normal/corrected-to-normal vision (iv) not colorblind (v) not easily susceptible to motion sickness and (vi) no COVID-19 symptoms. Note, two participants had an eight-day interaction gap, rather than seven, due to a campus wide closure for inclement weather (effect assumed to be negligible).

**SAE Level 2 Driving and Simulator**

Participants were tasked with monitoring a self-driving car that simulated SAE Level 2 driving automation in an urban area. The automated driving was simulated using the "Wizard of Oz" technique (Wang et al., 2017) by playing a pre-recorded drive while still granting participants the ability to take over control through braking. Once the AV was stopped and participants released the brake, the AV found and resumed driving the pre-recorded route via the technique "closest approximate simple curving" (Wu et al., 2020). The pre-recorded drives followed one of two routes, each of which spanned 15 blocks, had a balanced number of left and right turns, and included 10 intersections where the AV would stop. The study utilized a medium-fidelity driving simulator consisting of a steering wheel, pedals, and three 55" curved screens. The screens provided a field of view of ≈120°. All simulated environments were rendered using Unreal Engine 4.18. Half of the AV interactions were designed to have high automation transparency; therefore, AR cues were overlaid in 50% of the simulator trials via a software developed in Unity.

**Experiment Design**

The experiment is designed to measure participants' trust in an SAE Level 2 AV over the course of two interactions separated by an interaction gap of one week. A one-week gap is chosen to emulate intermittent uses of an SAV (e.g., airport to hotel and back, weekly errands). Each interaction (≈45 min) consists of five micro-interactions (a tutorial and four normal drives about 7 minutes each) separated by transition gaps (of 2-3 minutes) during which trust is measured as well as scene and automation variables altered. Figure 1 is a visualization of the drive order, including the interactions, the interaction gap, the micro-interactions, and the transition gaps.

Overall reliability (automation performance), task difficulty (weather), and risk (scene complexity) vary for each micro-interaction. Automation transparency (head-up display) varies between interactions. Each participant completed every drive listed in Table 1 in varying orders, blocking for transparency. In an attempt to eliminate additional ordering effects, each interaction uses a balanced Latin square of size four. A uniform number of participants per order was intended; however, due to excluding some participants, 3/8 orders had one more or one less participant than the other five.

Each drive, or micro-interaction, has 10 events in the form of four-way intersections. Two factors are varied for each intersection: automation reliability and risk. Reliability varies at each intersection based on the AV's braking behavior. Low reliability intersections entail an abrupt stop (deceleration < 25m from crosswalk) while high reliability intersections involve a smooth stop (deceleration > 60m from crosswalk). Risk varies randomly at each intersection with the absence or presence of pedestrians. Low risk intersections contain no pedestrians while high risk ones contain eight pedestrians (two per crosswalk). Five of the 10 intersections in each drive are randomly assigned to be high risk.

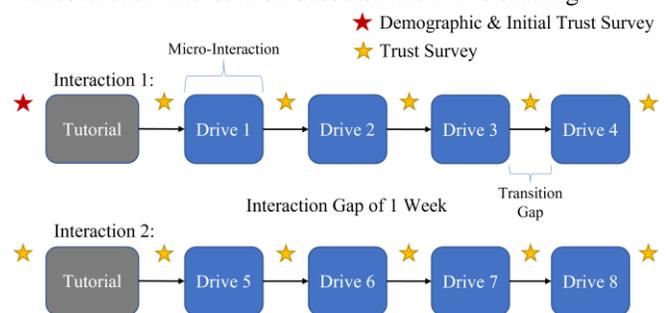

**Figure 1:** The drive order and sequence of events that participants engaged in while completing the study

Table 1: Variation of factors across the drives. Each drive includes five high risk intersections.

| Drive (Micro-Interaction) | Automation Transparency | Task Difficulty | Overall Reliability | # Low Reliability Intersections |
|---|---|---|---|---|
| A | High | Low | 100% | 0 |
| B | High | Low | 80% | 2 |
| C | High | High | 80% | 2 |
| D | High | High | 60% | 4 |
| E | Low | Low | 100% | 0 |
| F | Low | Low | 80% | 2 |
| G | Low | High | 80% | 2 |
| H | Low | High | 60% | 4 |

Task difficulty varies by the weather in the scene. High task difficulty includes fog and snow, limiting the visibility of road signs, pedestrians, and other objects in the scene. Low task difficulty conditions simulate sun with perfect visibility.

Overall reliability is correlated with task difficulty so that increased task difficulty results in lower overall reliability. Overall reliability has three levels: 100% reliable, 80% reliable, and 60% reliable where 10/10, 8/10, and 6/10 intersections are reliable, respectively. 100% reliable drives only occur with low task difficulty while 60% reliable drives only occur with high task difficulty. 80% reliable drives occur in an equal number of high and low task difficulty conditions. Thus, four combinations of overall reliability and task difficulty exist for each drive, as seen in Table 1.

Finally, automation transparency is varied between two levels: high and low. High transparency involves AR cues, visualized using bounding boxes for traffic signs, pedestrians, and cars, as well as path-prediction lines for pedestrians and cars (see Figure 2). Low transparency has no AR cues. Table 1 shows that the four combinations of overall reliability and task difficulty exist for high and low transparency, producing eight drives. Although all these variables affect human trust, this work only focuses on the effects of an interaction gap on trust.

**Experiment Procedure**

Participants were given instructions to monitor the driving behavior of a self-driving car, where the lane position, vehicle speed, braking, and all other driving functionality was controlled by the vehicle. If at any point they felt the driving was unreliable, they were instructed to bring the car to a complete stop. Once stopped, the self-driving car would resume driving. Participants completed an eight-intersection tutorial at the beginning of each interaction to practice braking and to become familiar with the user interface.

Throughout the experiment, participants completed questionnaires regarding their perceived *cumulative* trust of the AV after each drive. Cumulative trust refers to one's current trust of the AV based on *all* drives, not just the most recent. The survey consisted of 12 questions measured on a 7-point Likert scale adapted from (Jian et al., 2000) to be specific to the SAE Level 2 simulator vehicle. An example statement of the adapted survey is: *"The self-driving car is reliable."* Additionally, to ensure a uniform notion of trust, participants were given an adapted definition of trust from Lee and See (2004): *"Trust is defined as your attitude that the AV will help you achieve your goal of driving safely in a situation characterized by uncertainty and vulnerability."*

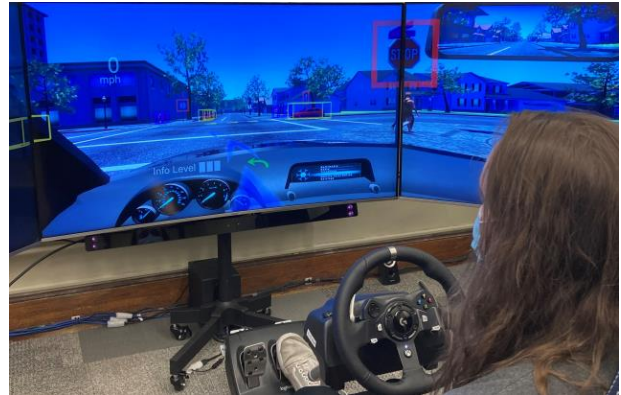

Figure 2: Participant completing a micro-interaction on high transparency.

**RESULTS**

**Changes in Trust During the Interaction Gap**

Inspired by the Ebbinghaus Forgetting Curve (Murre and Dros, 2015), we hypothesize that a participant's change in trust across the interaction gap is inversely related to their change in trust across the first interaction. Trust in this section refers to a participant's average trust score, the mean of their 12 responses to the items in the aforementioned questionnaire (Jian et al., 2000).

To test this hypothesis, the overall change in trust across the first interaction is correlated with the change in trust across the interaction gap. The *overall change in trust across the first interaction* is determined by subtracting each participant's trust after the tutorial from their trust at the end of drive 4. The *change in trust across the interaction gap* is determined by subtracting participants' trust at the end of drive 4 from participants' trust when they returned at the onset of the second interaction. The Pearson's product-moment correlation between the change in trust over the interaction gap and interaction 1 is negative, statistically significant, and moderate ($r = -0.57$, 95% CI [-0.79, -0.23], $t(23) = -3.35$, $p < .003$). The moderate correlation coefficient ($r = -0.57$) with its 95% confidence interval ([-0.79, -0.23]) provides compelling evidence that changes in trust across the interaction gap are inversely correlated to changes in trust across interaction 1.

Seventeen participants' trust increased, one participant's trust remained the same, and seven participants' trust decreased during interaction 1. Observing these participants across the gap: (i) 12/17 of those whose trust increased during interaction 1 decreased during the interaction gap (ii) 1/1 of those whose trust remained the same during interaction 1 increased during the interaction gap (iii) 6/7 of those whose trust decreased during interaction 1 increased during the interaction gap. Thus, 18 of the 25 participants performed according to our hypothesis that trust will increase or decrease over an interaction gap, depending upon if trust decreases or increases, respectively, during interaction 1 (see Figure 3).

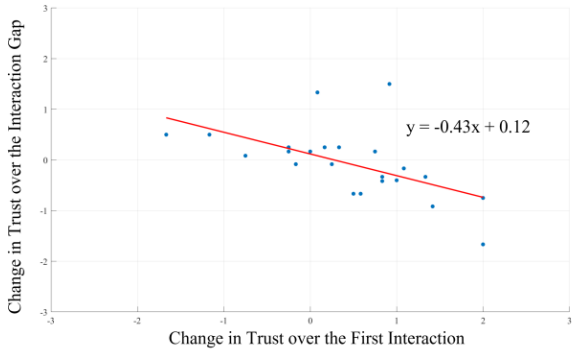

**Figure 3:** Inverse trust correlation between the first interaction and the interaction gap is shown by a scatter plot and linear line of best fit.

**Aspects of Trust that Change During the Interaction Gap**

Each phrase in the empirical, 12-item questionnaire developed by Jian et al. (2000) relates to a different aspect of trust or distrust. Five items relate to distrust while seven relate to trust; for simplicity, both trusting and distrusting items will be referred to as *aspects* of trust. This definition agrees with Jian et al.'s (2000) finding that trust and distrust are indeed opposites and not different concepts. The trust described in the prior section is an average of the 12-item questionnaire which gives a strong measure of overall trust. However, to perceive which aspects of trust are most affected by an interaction gap, this section regards each trust item individually.

The change in magnitude for each aspect of trust, across the first interaction and then across the interaction gap, is determined. Then, a Pearson correlation test is performed on these changes in trust for each of the 12 aspects of trust. Table 2 provides the correlation coefficient, confidence interval, t-value, p-value, strength, direction, and significance for each aspect of trust. Strength is evaluated on a 3-point scale with the three points being: weak ($0 \leq r < .33$), moderate ($.33 \leq r < .67$), and strong ($.67 \leq r \leq 1$). Schober et al. (2018) recommend exercising caution when oversimplifying strengths of correlations on scales such as these; thus, the strengths in Table 2 are starting points for interpreting the correlation.

The interaction gap and the first interaction are negatively correlated for every aspect of trust; however, four of the twelve aspects of trust (*deceptive, underhanded, harmful, and dependable*) are not statistically significant. *Deceptive* arguably has no correlation ($r = -0.09$) and the greatest p-value ($p = 0.665$). The remaining eight aspects of trust have either a moderate or strong correlation, with *suspicious* as the only aspect with a strong correlation ($r = -.80$). *Integrity, reliable, and trust* all have nearly identical correlation coefficients ($r = -0.52, -0.55, -0.52$) and confidence intervals ([-0.76, -0.15], [-0.78, -0.20], [-0.76, -0.15]).

## DISCUSSION

**Trust During the Interaction Gap**

The fact that the Pearson's product-moment correlation between $\Delta T_{int,1}$ and $\Delta T_{gap}$ is negative, statistically significant, and moderate ($r = -0.57$, 95% CI [-0.79, -0.23], $t(23) = -3.35$, $p < .003$) suggests that people "forget" some of their gained trust or distrust. During periods of no interaction with automation (interaction gaps), the effects of trust building or reducing experiences from the previous interaction begin to fade. Figure 3 shows the inverse correlation between the first interaction and the interaction gap. A slope $m = -0.57$ implies that trust in the SAE Level 2 AV increases or decreases more during the first interaction than it decreases or increases during the interaction gap, respectively. Therefore, changes in trust over the interaction gap typically revert *toward* participants' initial trust in the automation.

**Table 2:** Individual Jian et al. (200) Survey Factors - Correlating the Interaction Gap to Interaction 1

| Aspect of Trust | r | 95% CI | t-value | p-value | Strength | Direction | Sig. |
|---|---|---|---|---|---|---|---|
| Deceptive | -0.09 | [-0.47, 0.32] | -0.44 | 0.665 | Weak | Neg. | No |
| Underhanded | -0.27 | [-0.60, 0.14] | -1.35 | 0.190 | Weak | Neg. | No |
| Suspicious | -0.80 | [-0.91, -0.59] | -6.35 | <.001 | Strong | Neg. | Yes |
| Wary | -0.61 | [-0.81, -0.28] | -3.69 | 0.001 | Moderate | Neg. | Yes |
| Harmful | -0.37 | [-0.67, 0.02] | -1.94 | 0.065 | Moderate | Neg. | No |
| Confident | -0.48 | [-0.74, -0.11] | -2.65 | 0.014 | Moderate | Neg. | Yes |
| Security | -0.40 | [-0.69, 0] | -2.10 | 0.047 | Moderate | Neg. | Yes |
| Integrity | -0.52 | [-0.76, -0.15] | -2.89 | 0.008 | Moderate | Neg. | Yes |
| Dependable | -0.26 | [-0.59, 0.15] | -1.28 | 0.214 | Weak | Neg. | No |
| Reliable | -0.55 | [-0.78, -0.20] | -3.15 | 0.005 | Moderate | Neg. | Yes |
| Trust | -0.52 | [-0.76, -0.15] | -2.90 | 0.008 | Moderate | Neg. | Yes |
| Familiar | -0.57 | [-0.79, -0.23] | -3.33 | 0.003 | Moderate | Neg. | Yes |

Interestingly, both increases in trust (positive first interaction) and decreases in trust (negative first interaction) with the SAE level 2 AV were mitigated after the interaction gap. This means that participants whose trust in the AV increased while interacting with it were not as trusting of it a week later. Additionally, those whose trust in the AV diminished while interacting with it were not as *distrusting* of it a week later. In other words, humans tend to revert toward their initial perceptions of, or even their dispositions to trust, automated mobility. As SAVs are introduced in coming years, researchers and designers should strive to deliver poignant experiences so trust can be readily retained.

**Aspects of Trust During the Interaction Gap**

*Perceptions of Deception, Underhandedness, and Harmfulness Remain or Increase Over the Interaction Gap.* There are no significant correlations between the interaction gap and interaction 1 for neither deceptive, underhanded, nor harmful perceptions. In fact, 92%, 88%, and 80% of participants found the AV to be equally or more deceptive, underhanded, and harmful after the interaction gap, respectively. Therefore, it can be inferred that a one-week interaction gap rarely reduces perceptions of deception, underhandedness, or harmfulness; rather, users will perceive the AV to have the same, or even an increased, tendency to deceive and/or harm.

*Suspicions Either Don't Change or Change Inversely Over the Interaction Gap.* Eighty-four percent of participants found the AV to be equally or more suspicious after the gap. Every participant whose suspicions drastically changed (n=7) during interaction 1 had their suspicions revert to or toward their initial suspicion after the gap, supporting the high inverse correlation ($r = -0.81$). These results suggest that suspicions rarely decrease during a gap except when suspicions were significantly heightened during a prior interaction.

*Wariness Either Doesn't Change or Changes Inversely Over the Interaction Gap.* Similar to the other four aspects of distrust, 76% of participants were equally or more wary of the AV after the interaction gap. However, nearly half of the participants whose trust changed during interaction 1 had their trust revert over the interaction gap, thereby supporting the moderate correlation ($r = -0.61$).

*Confidence Holds Steady or Increases over the Interaction Gap.* Eighty percent of participants were equally or more confident in the AV after the gap. This appears at odds with the five aspects of distrust, signifying confidence is interpreted differently and is indeed a distinct aspect of trust.

*Security Primarily Holds Steady over the Interaction Gap.* Sixty percent of participants had no change in perceived security of the AV over the gap while only 24% had an inverse change in trust during the interaction gap.

*Dependability Primarily Holds Steady over the Interaction Gap.* Similar to perceived security, 60% of participants had no change in perceived dependability of the AV over the gap.

*Integrity, Trust and, Reliability Perform Similarly Over the Interaction Gap.* In this context, integrity, trust, and reliability refer to participants' *perceived integrity, perceived trust*, and *perceived reliability* of the AV. The three aspects nearly all have the same correlation coefficient ($r = -0.52, -.52, -0.55$) on almost identical confidence intervals. Over the interaction gap, integrity, trust, and reliability had a similar number of participants whose trust increased (n=9,8,7) and decreased (n=7,6,5), respectively. These results suggest that integrity, trust, and reliability are closely related aspects of trust and are "forgotten" similarly over an interaction gap.

*Familiarity Remains Constant or Decreases Over the Interaction Gap.* For 96% of participants, their familiarity with the AV increased or remained constant (with respect to the tutorial) over interaction 1. Then 88% of participants indicated that their familiarity with the AV decreased or stayed constant over the gap. Thus, the inverse correlation ($r = -0.57$) between the gap and interaction 1 mainly applies to an *increase* during interaction 1 and a *decrease* over the gap. Consequently, this signifies that gains in familiarity are strong in dynamic learned trust while weaker in initial learned trust.

## CONCLUSION

This work is a foundational contribution toward an understanding of how trust changes during time spans of no interaction with an automation. We show that an inverse correlation exists between changes in trust during an interaction gap and an initial interaction. Additionally, we determined that specific aspects of trust in automation are affected differently over an interaction gap. Future work includes identifying which aspects of trust correspond to design features in autonomous vehicles as well as examining multiple interaction gaps and gaps of varying lengths.